\documentclass[journal]{IEEEtran}

\ifCLASSINFOpdf
\usepackage[pdftex]{graphicx}
\DeclareGraphicsExtensions{.pdf,.jpeg,.png}
\else
\usepackage[dvips]{graphicx}
\DeclareGraphicsExtensions{.eps}
\fi

\usepackage{comment}
\usepackage{url}
\usepackage{hyperref}
\usepackage{breakurl}

\begin{document}
\title{A Design of Scintillator Tiles Read Out by Surface-Mounted SiPMs for a Future Hadron Calorimeter}
%
%

\author{Yong Liu, Volker B{\"u}scher, Julien Caudron, Phi Chau, Sascha Krause, Lucia Masetti, \\
        Ulrich Sch{\"a}fer, Rouven Spreckels, Stefan Tapprogge, Rainer Wanke
\thanks{Manuscript received November 26, 2014. This work was supported by the Cluster of Excellence ``Precision Physics, Fundamental Interactions, and Structure of Matter'' (PRISMA) funded by the German Research Foundation (DFG) within the German Excellence Initiative.}
\thanks{Yong Liu, Volker B{\"u}scher, Julien Caudron, Phi Chau, Sascha Krause, Lucia Masetti, Ulrich Sch{\"a}fer, Rouven Spreckels, Stefan Tapprogge and Rainer Wanke are with Institut f{\"u}r Physik and PRISMA Detector Lab, Johannes Gutenberg-Universit{\"a}t Mainz, D-55099 Mainz, Germany.}%
\thanks{The corresponding author's e-mail: yong.liu@uni-mainz.de.}%

}

\maketitle
\pagestyle{empty}
\thispagestyle{empty}

\begin{abstract}
Precision calorimetry using highly granular sampling calorimeters is being developed based on the particle flow concept within the CALICE collaboration. One design option of a hadron calorimeter is based on silicon photomultipliers (SiPMs) to detect photons generated in plastic scintillator tiles. Driven by the need of automated mass assembly of around ten million channels stringently required by the high granularity, we developed a design of scintillator tiles directly coupled with surface-mounted SiPMs. A cavity is created in the center of the bottom surface of each tile to provide enough room for the whole SiPM package and to improve collection of the light produced by incident particles penetrating the tile at different positions. The cavity design has been optimized using a GEANT4-based full simulation model to achieve a high response to a Minimum Ionizing Particles (MIP) and also good spatial uniformity. The single-MIP response for scintillator tiles with an optimized cavity design has been measured using cosmic rays, which shows that a SiPM with a sensitive area of only $\mathbf{1\times1~mm^2}$ (Hamamatsu MPPC S12571-025P) reaches a mean response of more than 23 photon equivalents with a dynamic range of many tens of MIPs. A recent uniformity measurement for the same tile design is performed by scanning the tile area using focused electrons from a $\mathbf{^{90}Sr}$ source, which shows that around 97\% (80\%) of the tile area is within 90\% (95\%) response uniformity. This optimized design is well beyond the requirements for a precision hadron calorimeter.
\end{abstract}


\section{Introduction}
%
%
%
%
\IEEEPARstart{T}{}{he} precision physics at future linear collider experiments including Standard Model measurements and searches for new physics beyond the Standard Model requires unprecedented reconstruction capabilities of di-jet and multi-jet final states. The desired jet energy resolution $\sigma_{E}/E=30\%/\sqrt{E}$ can be achieved by particle flow algorithms, which will in turn require highly granular calorimetry with all read-out electronics fully integrated into the detector volume~\cite{AHCAL_2013}. One option of highly granular hadron calorimeters developed within the CALICE collaboration uses small and thin plastic scintillator tiles individually read out by silicon photomultipliers (SiPMs) \cite{AHCAL_PhysPrototype}. As high granularity will lead to a large number of channels, the implementation of around ten millions scintillator tiles and SiPMs in the final hadron calorimeter is challenging and can only be feasible via automated mass assembly. Surface-mounted SiPMs (SMD-SiPMs) are an ideal candidate for this goal since they are more tolerant in mechanical positioning than SiPMs with pins (THT-SiPMs, where THT stands for Through-Hole Technology). We have also observed that part of a scintillator tile was melted during machine soldering of a THT-SiPM which was beforehand glued into the side surface of the tile~\cite{Phi_MA}, which also makes the mass assembly of THT-SiPMs more challenging. 

But directly coupling an SMD-SiPM to an intact scintillator tile does not result in an adequately uniform spatial response. There have been efforts of shaping a cavity in a scintillator tile to improve the response uniformity \cite{NIU_MegaTile1, NIU_MegaTile2}. We have studied further on how the geometry and surface properties of a cavity influences light collection and developed a dedicated cavity design for a given SiPM package based on the GEANT4 full simulation. In our design, an SMD-SiPM can be directly soldered onto a readout Printed Circuit Board (PCB) and fully placed inside a cavity. Therefore, all SiPMs can be soldered in the same way as other standard SMD-components for the PCB. After the mass soldering process, tiles can then be directly placed on top of SiPMs individually, which would be feasible via automated mass assembly. Such a design aims to achieve both high response to incident particles and good spatial response uniformity.

Herein we first present simulation results of an optimized cavity design, followed by cosmic-ray measurements for the response of a scintillator tile using the same cavity design to Minimum Ionizing Particles (MIPs). Finally response uniformity measurements for the cavity design using a radioactive source to scan the whole tile area are summarized.

\section{Simulation studies for the cavity optimization}
In order to study and optimize the cavity design, a GEANT4-based full simulation model has been built, including all related physics process, such as scintillation, transportation and optical boundary processes. Detailed geometry descriptions of a scintillator tile with a cavity and a SiPM package have been implemented. The emission spectrum of a chosen scintillator (BC408) as well as the photon detection efficiency (PDE) of a SiPM candidate (Hamamatsu MPPC S12571-025P) at different photon wavelengths are interpolated based on data sheets respectively. The wavelength-dependent reflectivity of two types of 3M reflective foil (ESR and DF2000MA) wrapped around the tile is also taken into account, based on two independent measurements~\cite{Refl1, Refl2}. 

\begin{figure}[!t]
\centering
\includegraphics[width=3.5in]{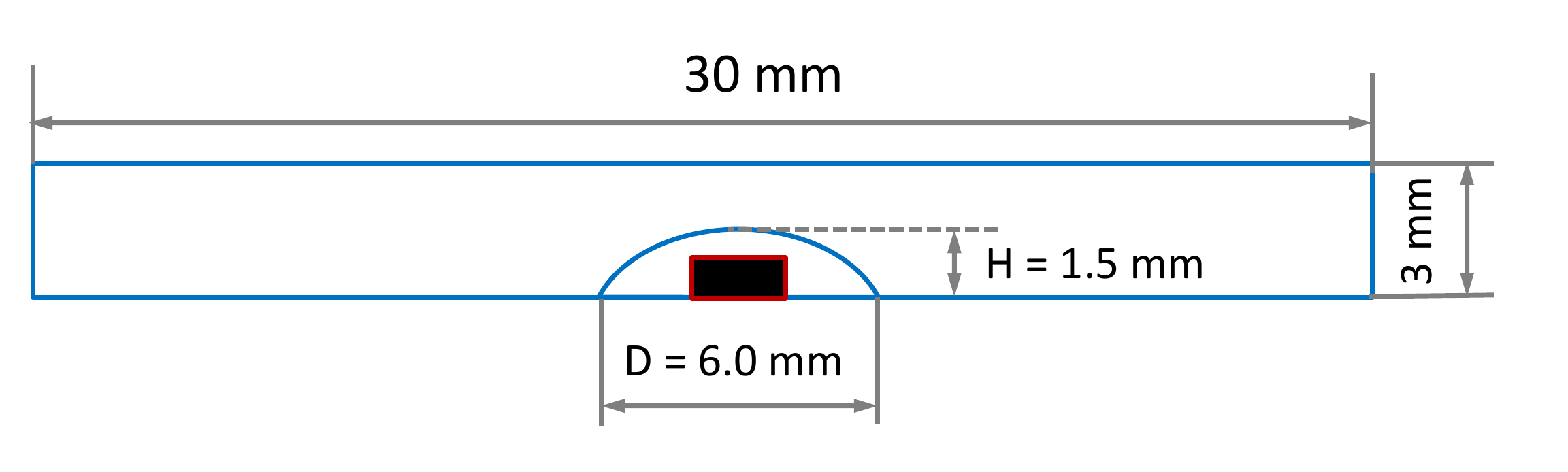}
\caption{Schematics of a scintillator tile with an optimized dome-shaped cavity (H and D are the height and diameter of the dome respectively) and an SMD-SiPM (red) completely inside the cavity}
\label{fig_Cavity}
\end{figure}

The size of each tile is $\mathrm{30\times30\times3~mm^3}$, as required by the aforementioned high granularity. In the simulation, placing a cavity in the center of the tile surface shows better response uniformity than along a side or in a corner. Among several different shapes, a dome-shaped cavity is chosen, which shows promising response uniformity and simplifies shaping of the cavity via mechanical drilling and polishing. Different geometry parameters of the dome (i.e. dome height and dome diameter on the bottom surface) and surface properties for both the tile and the cavity (i.e. polished or ground) have been varied in the simulation, which turns out that polished surfaces of the tile and the cavity can improve response uniformity. Geometry parameters of the dome have been chosen as shown in Fig.~\ref{fig_Cavity}, which ensures enough space for the SiPM package without degrading response uniformity.

\begin{figure}[!t]
\centering
\includegraphics[width=3.5in]{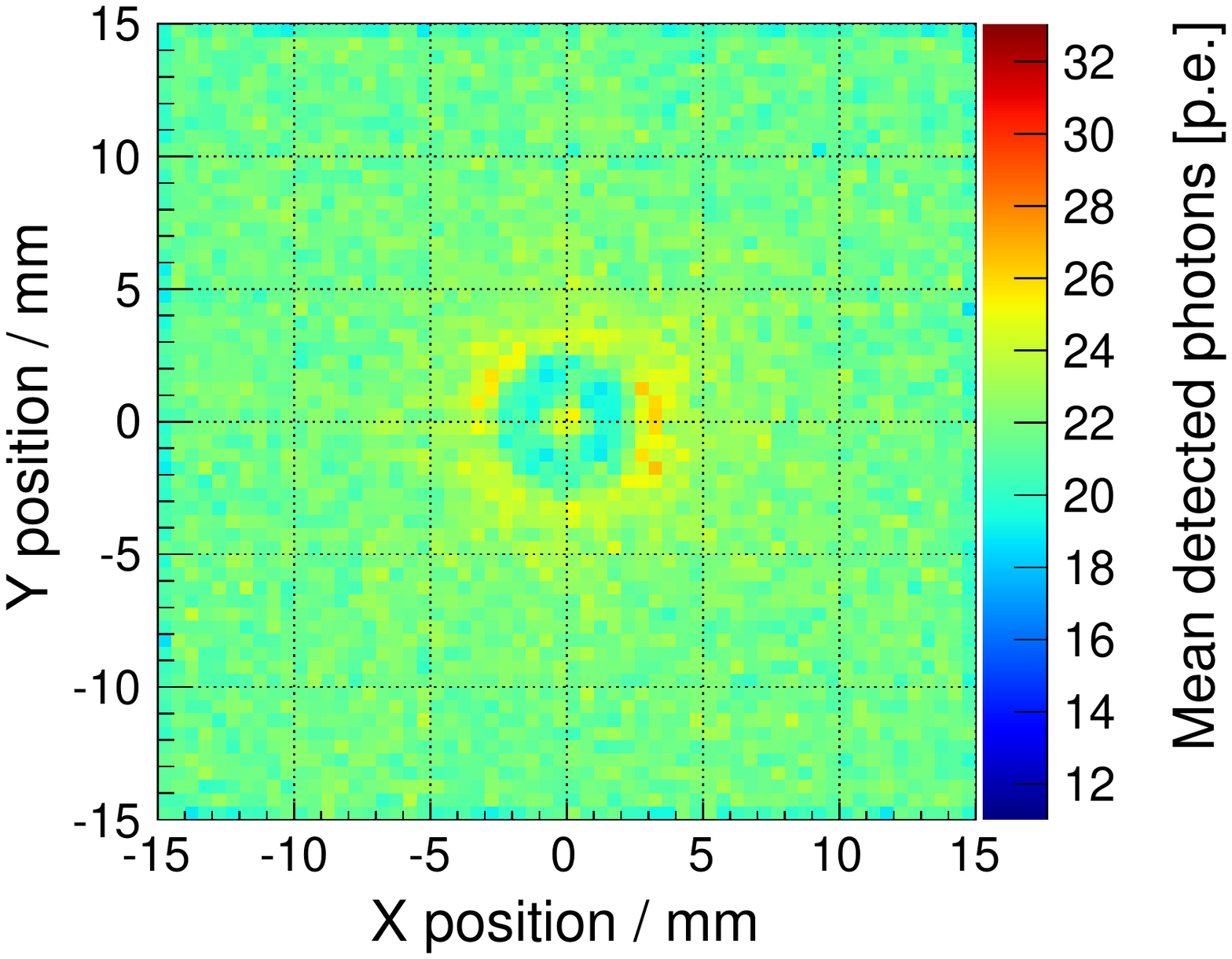}
\caption{Simulated uniformity using the optimized cavity design with 2.28 MeV electrons perpendicularly penetrating the tile}
\label{fig_Uniformity_Sim}
\end{figure}

In order to be comparable with later uniformity measurements using a beta source, low-energy electrons (2.28 MeV) have been used in the simulation to scan the tile area ($\mathrm{30\times30~mm^2}$) in a $\mathrm{0.5\times0.5~mm^2}$ step size. A two-dimensional histogram of simulated response uniformity is shown in Fig.~\ref{fig_Uniformity_Sim}, where a single colorful point represents the mean response for electrons passing through this area. The global mean response across the tile area is around 22.0 p.e. (photon equivalents) and 97.1\% (80.8\%) of the tile area is within 10\% (5\%) deviation from the mean value.

\section{Cosmic-ray measurements}
We have constructed in Mainz a customized cosmic-ray test setup for measuring responses of scintillator tiles to muons, which are ideal candidates of MIPs. One of tiles under test is placed between two trigger scintillators coupled to two PMTs by light guides. The coincidence of two trigger signals ensures a muon track passing though the tile and also constrains the incident angle of each muon.

\begin{figure}[!t]
\centering
\includegraphics[width=3.5in]{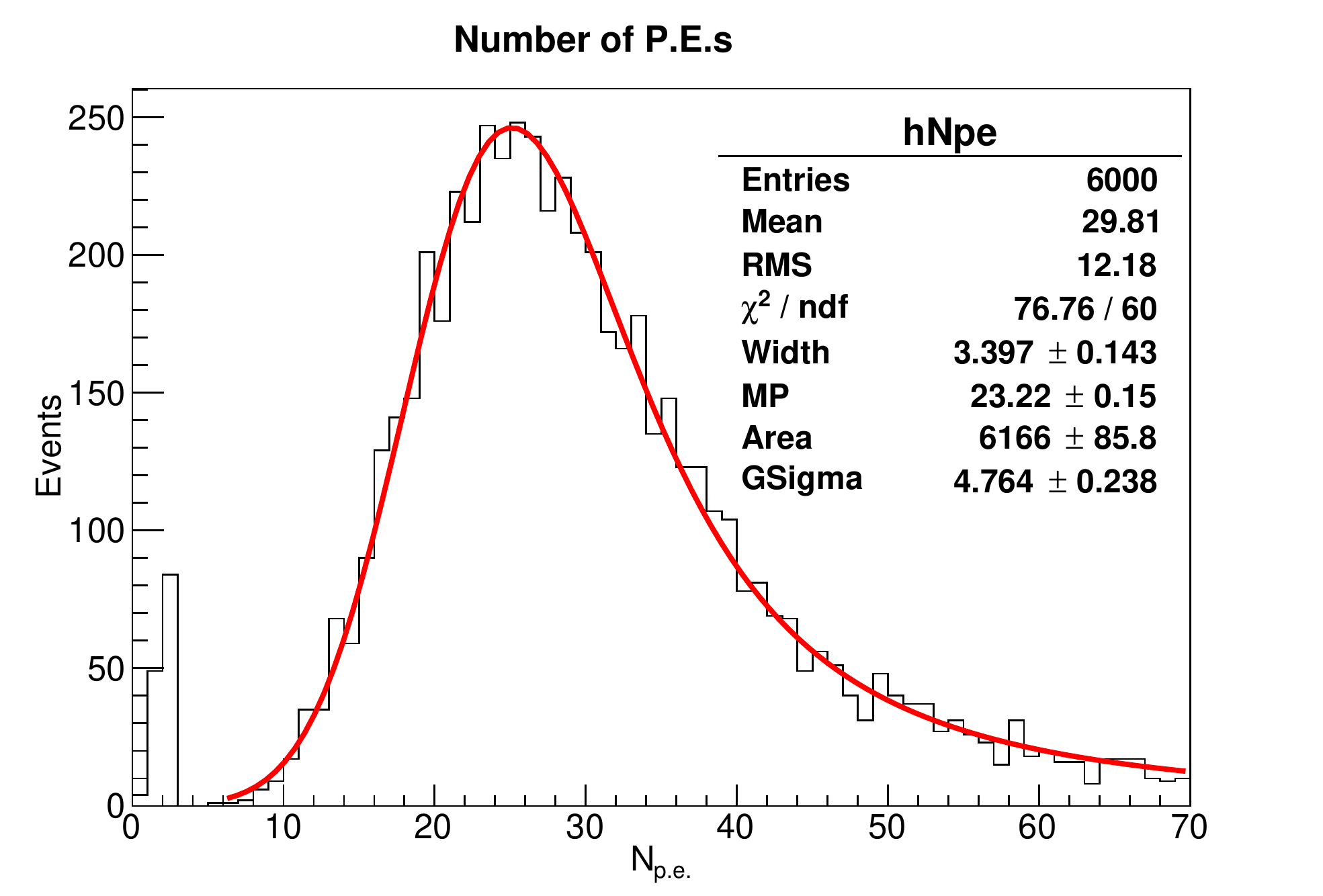}
\caption{Cosmic-ray measurement: single-MIP responses of a scintillator tile with a Hamamatsu SMD-SiPM coupled inside the optimized cavity}
\label{fig_1MIP_MZ}
\end{figure}

\begin{figure}[!t]
\centering
\includegraphics[width=3.5in]{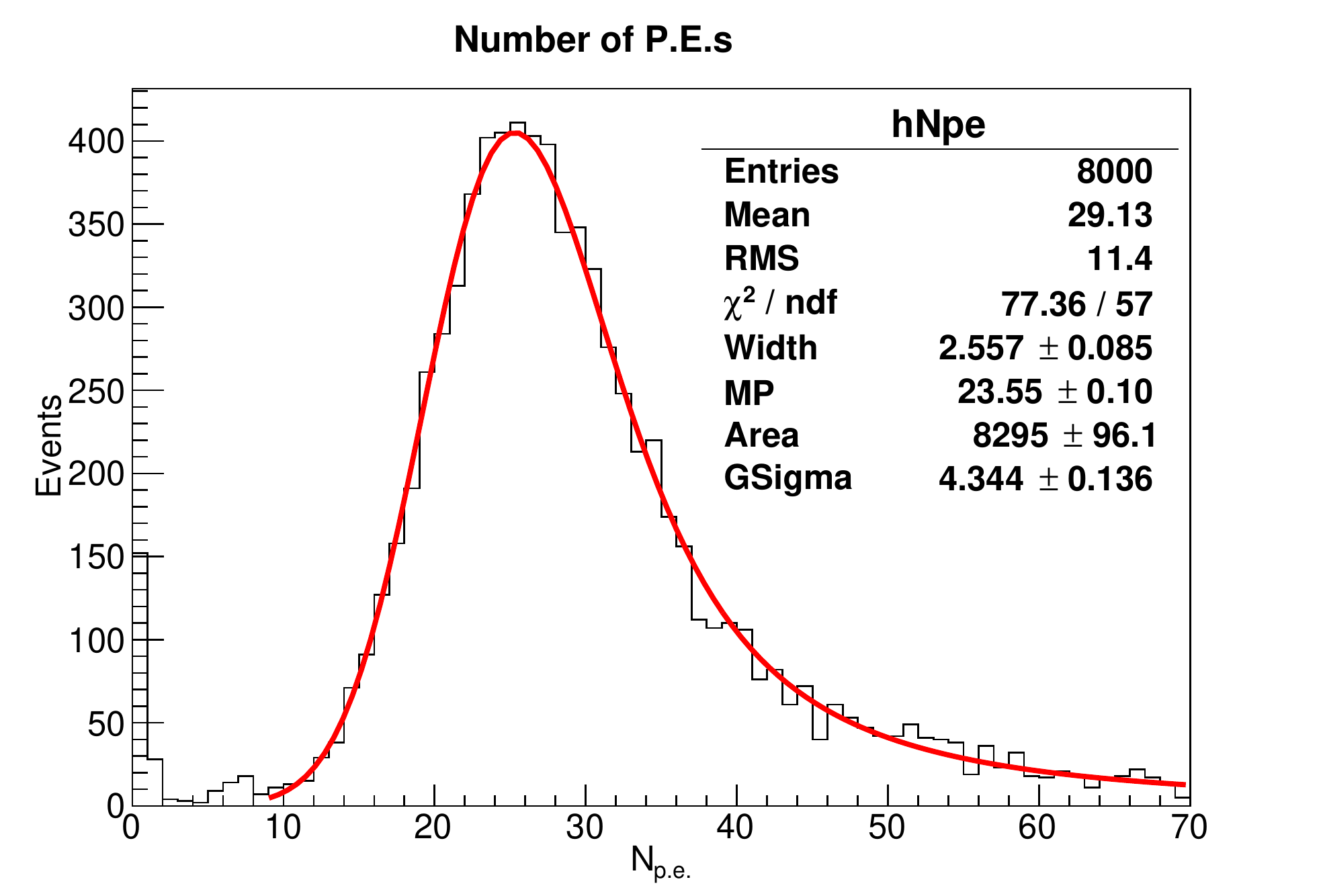}
\caption{Cosmic-ray measurement: single-MIP responses of a scintillator tile with a KETEK SiPM coupled onto the side surface}
\label{fig_1MIP_HH}
\end{figure}

The SMD-SiPM module MPPC S12571-025P has a sensitive area of $\mathrm{1\times1~mm^2}$ with $\mathrm{25~\mu m}$ pixel pitch. One such SMD-SiPM is soldered onto a readout PCB and a scintillator tile is fixed on top of the PCB. With the optimized cavity design, this SiPM detected 23.2 p.e. on average to a muon, as shown in Fig.~\ref{fig_1MIP_MZ}. As a reference, a tile using the side-surface design~\cite{Hamburg_Tile} is also measured in this setup using a SiPM KETEK PM1125NT, which shows a mean single-MIP response of 23.5 p.e., as shown in Fig.~\ref{fig_1MIP_HH}. 

Tab.~\ref{Table_CosmicRay} lists mean single-MIP responses to muons of different configurations, where the mean $N_{p.e.}$ is obtained by fitting a SiPM charge spectrum using a Landau convoluted with Gaussian distribution. The gain used in the fitting was extracted from SiPM dark count spectra. Two SiPMs were both operated in the same environment (temperature controlled to be $\mathrm{21.2\pm0.5^{\circ}C}$) and at suggested overvoltages (around 3.0~V above the breakdown voltage respectively). 

The optimized cavity design using a relatively smaller sensitive area can reach a similar single-MIP response as the side surface design. This single-MIP response is high enough to suppress the SiPM dark counts with negligible efficiency loss. With all 1600 pixels in the SMD-SiPM, it can also detect many tens of MIPs which ensures a large dynamic range.


\begin{table} \centering
\caption{Cosmic-ray measurements for scintillator tiles}
\label{Table_CosmicRay}
\renewcommand{\arraystretch}{1.2}
\begin{tabular}{cccc}
\hline
Design & SiPM Type & Sensitive Area & Mean $N_{p.e.}$ \\
\hline
SMD design & MPPC S12571-025P & $\mathrm{1.0\times1.0~mm^2}$ & 23.2\\
Side-surface design & KETEK PM1125NT & $\mathrm{1.2\times1.2~mm^2}$ & 23.5\\

\end{tabular}
\end{table}

\section{Uniformity measurements}
The response uniformity of an SMD-SiPM inside the cavity of a scintillator tile has been measured in a dedicatedly designed test stand in the Max Planck Institute for Physics in Munich with a $\mathrm{^{90}Sr}$ source \cite{MPI_Scan}. The beta source and a cubic trigger scintillator, which were mounted on a high-precision motorized translation stage, can be finely moved together horizontally. A scintillator tile under study was fixed and placed between the source and the trigger scintillator. The trigger cube, which is read out by a SiPM, helps to select a fraction of electrons with higher energy, which can completely pass through the tile under study. More than 6 p.e. were required to be detected at the trigger SiPM during measurements. Uniformity scans were accomplished by horizontally translating the source in a fine step size of $\mathrm{0.5\times0.5~mm^2}$.

\begin{figure}[!t]
\centering
\includegraphics[width=3.5in]{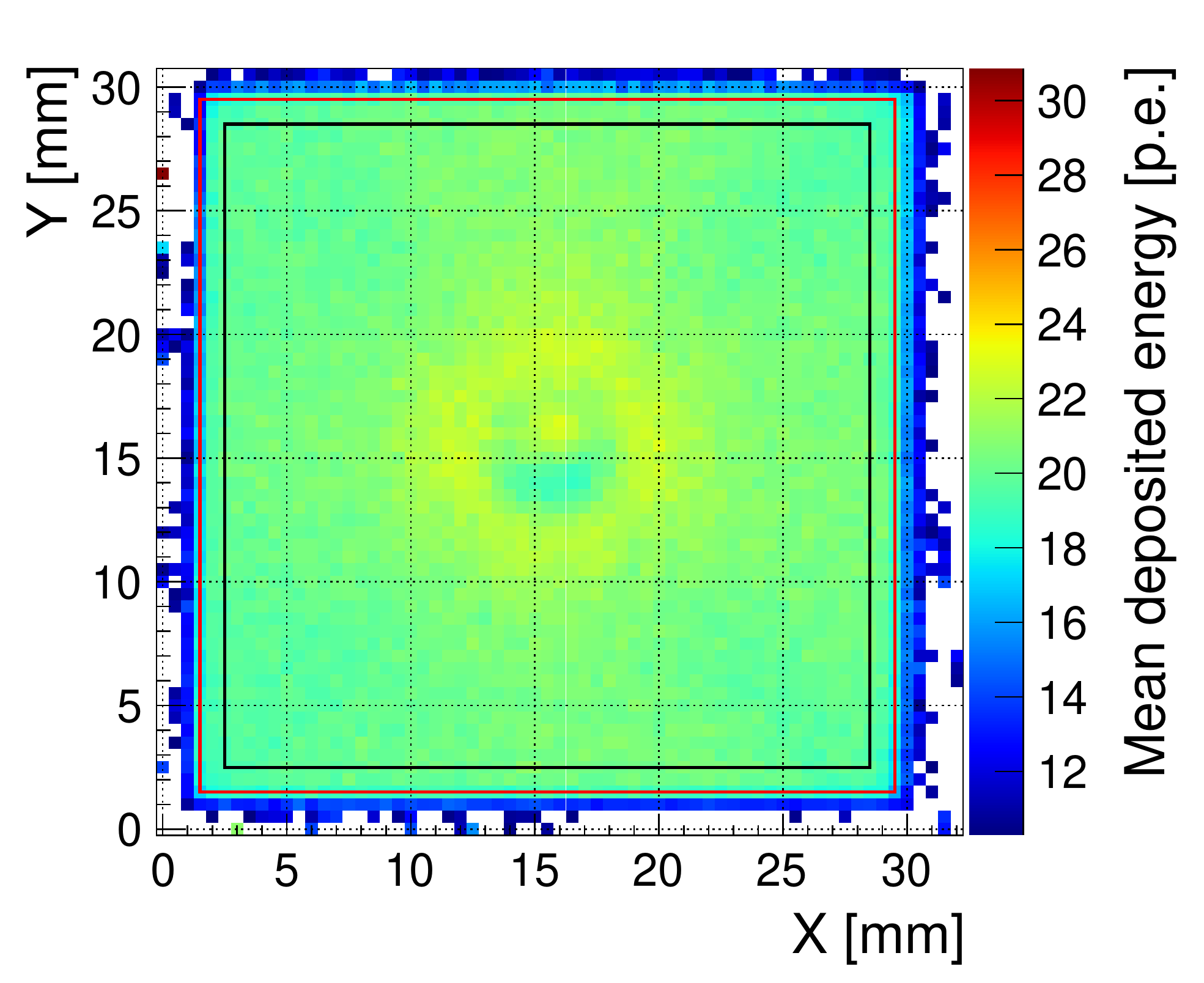}
\caption{A 2D uniformity scan of a scintillator tile with the optimized cavity design, where the black square represents the area to determine the global average response across the tile (20.6 p.e.), and the red square is used to calculate areal fractions within certain deviations from this global mean response.}
\label{fig_UniformityMean}
\end{figure}

Results of one uniformity scan for the optimized cavity design are shown in Fig.~\ref{fig_UniformityMean}, which show around 97.1\% (80.8\%) of the tile area is within 10\% (5\%) deviation from the average response of 20.6 p.e. to electrons. The blue ring pattern in the center of the uniformity map has been studied in the simulation; it is due to the combination effects of the SiPM non-sensitive package and a round hole cut in the center of the reflective foil for the SiPM package, where relatively more photons are absorbed.

The uniformity measurement shows excellent response uniformity of the optimized cavity design and it is consistent with the prediction from the Geant4-based full simulation model.

\section{Summary and Outlook}
Based on GEANT4 full simulation studies, we have developed a cavity design for scintillator tiles which ensures high light collection efficiency and uniform response across the tile area. Cosmic-ray and uniformity measurements performed afterwards show an adequately high single-MIP response, a large dynamic range to detect many tens of MIPs and excellent spatial response uniformity, which is consistent with the full simulation. In this design, the SMD-SiPM package (Hamamatsu MPPC-S12571-025P) can be directly soldered onto a readout PCB and be fully placed inside the cavity, which is suitable for the automated mass assembly of a future precision hadron calorimeter.

The design and assembly of a 144-channel hadron calorimeter base unit (SMD-HBU) based on this design have been finished successfully. We plan to test this SMD-HBU extensively in an upgraded cosmic-ray test stand and in beam facilities to evaluate its performance.

Lately a new low-crosstalk SiPM prototype from Hamamatsu has been characterized at a stable temperature ($\mathrm{21.0^{\circ}C}$). Preliminary results show that the typical dark count rate (DCR) is reduced significantly (33 kHz) and the typical crosstalk probability is low ($<1$\%) due to the trench design. As a comparison,  the current module (Hamamatsu MPPC-S12571-025P) has a typical DCR of 100 kHz and its cross-talk probability is much higher (23\%). This new SiPM would be a promising photosensor candidate for the future scintillator-based hadron calorimeter.


\section*{Acknowledgments}
This work was carried out within the framework of the PRISMA Detector Lab at the University of Mainz. The authors would like to thank the FLC group at DESY for the joint efforts of the SMD-HBU board design, the group of Prof. Dr. Erika Garutti at the University of Hamburg for the help of the foil cutting and tile wrapping as well as the group of Dr. Frank Simon at the Max-Planck Institute for Physics in Munich for the efforts of uniformity measurements. Yong Liu is also grateful to the kind communications and beneficial discussions on the SiPM with Hamamatsu GmbH and KETEK GmbH.



\bibliographystyle{IEEEtran}
%

\end{document}